\documentclass[aps,pre,twocolumn]{revtex4} 
\usepackage{hyperref,natbib,doi}
\usepackage{graphicx}
\usepackage{amsmath}
\usepackage{dcolumn}
\usepackage[utf8]{inputenc}
\usepackage{amssymb}
\begin{document}
\title{Electron plasma wake field acceleration in solar coronal and chromospheric plasmas}
\author{David Tsiklauri}
\affiliation{School of Physics and Astronomy, Queen Mary University of London, London, E1 4NS, United Kingdom}
\begin{abstract}
Three dimensional, particle-in-cell, fully electromagnetic simulations of 
electron plasma wake field acceleration applicable to solar atmosphere are presented. 
It is established that injecting driving and trailing electron bunches
into solar coronal and chromospheric plasmas,
results in electric fields ($-(20-5) \times 10^{6}$ V/m), 
leading to acceleration of the trailing bunch up to 52 MeV, starting from initial 36 MeV.
The results provide one of potentially important mechanisms for the 
extreme energetic solar flare electrons, invoking plasma wake field acceleration.
\end{abstract}

\maketitle

\section{Introduction}

Large solar flares can  accelerate particles which travel into interplanetary
space and also smash into the Sun producing X-ray, gamma-rays and cause
Earth ionospheric response.
Detailed theoretical models exist,  which consider various aspects of the generation of
x-rays in solar flares \citep{mb76}.
Occasionally ions are accelerated to few GeV and electrons to in excess of 100 MeV 
\cite{1988SoPh..115..313S}. 
There are also some extreme energetic solar flare events, the so-called 
electron-dominated flares, see e.g. section 14.2.1 from \citet{2005a},
where upper energy cutoff provides a true lower limit
of the maximum energy of accelerated electrons. In some cases the
maximum electron energy can be as high as $50$ MeV or more. Such high energies pose
a problem for conventional electron acceleration mechanisms.
\cite{1999ApJS..120..409V} routinely measure (for the 185 flare events) 
electrons with $10-25$ MeV
energies in excess of composite photon spectral model.
\citet{1989ApJ...346..523M} present 55 event survey of energy 
spectra of $0.1-100$ MeV interplanetary electrons 
originating from solar flares as measured by two spectrometers 
on board the ISEE 3 (ICE) spacecraft.
In this work we apply a novel mechanism, the electron plasma wake field acceleration,
to solar coronal and chromospheric plasma extreme energetic solar flare events.

The basic concepts of plasma acceleration 
based laser wake field acceleration 
were originally conceived by Tajima and Dawson \citep{td79}. 
Initial experiments for the plasma wake field were implemented 
by Joshi \citep{joshi}.
Usually a distinction is made what creates the plasma wake:
a laser or a charged particle beam. The latter case is
referred to as plasma wake field acceleration (PWFA), while
the former as laser wake field acceleration (LWFA).
Current experimental devices show accelerating gradients several 
orders of magnitude better (10s of GeV m$^{-1}$) than current 
RF-based conventional particle accelerators (10s of MeV m$^{-1}$).
\citet{l14}
made a significant progress in PWFA. 
In their plasma wake field 
accelerator, the plasma wave is created by a 20-GeV electron 
bunch from SLAC's linac. A second bunch of equally energetic 
electrons follows close behind. With SLAC's purpose-built 
Facility for Advanced Accelerator Experimental Tests (FACET) \cite{facet}, 
authors could place the trailing bunch at just the right spot in the plasma wave to 
increase the bunch energy by 1.6 GeV over just 30 cm of plasma.

Also there has been interesting progress in applying PWFA concepts to
astrophysical plasmas. This includes
astrophysical ZeV acceleration in the 
relativistic jet from an accreting supermassive blackhole \cite{Ebisuzaki20149,
Ebisuzaki2014}, focusing on ponderomotive acceleration by relativistic waves
\cite{lau15} and elecromagnetic aspects \cite{farinella16}.
A comprehensive, recent review is available \cite{tajima17}.

In a recent work of \citet{pt14} a beam of accelerated electrons was injected into a magnetized, 
Maxwellian, homogeneous, and inhomogeneous background plasma. It was 
established that in the case of increasing density along the path of an electron 
beam wave-particle resonant interaction of Langmuir waves (the same type of wave as in 
plasma wake field acceleration) 
with the beam electrons leads to an efficient particle acceleration.  
This is due to Langmuir waves drift to smaller wave-numbers, $k$, allowing them to 
increase their phase speed, $V_{ph}=\omega/k$, and, therefore, 
being subject to absorption by faster electrons. 

This novel aspect electron acceleration has been 
explored in the plasma wake field acceleration context by \citet{me2016}.
Thus, yet another motivation for the present work is to extend results of \citet{me2016}
to tens of MeV range of electron energies and study the possibility of
electron acceleration in the context of extreme solar flares. 

In what follows a brief overview of solar atmosphere 
physical parameters of relevance to this work is presented.
The observational constraint on non-thermal electron densities 
in the chromosphere is the non-thermal hard X-ray (HXR) flux. 
However this can be interpreted in at least two different ways:

1) The chromospheric HXR flux can be expressed as a non-thermal emission measure, see e.g. 
\citet{2009A&A...508..993B}, their equations 2 and 3. 
Here $EM_{nt}$ tells us the {\it instantaneous} 
number density of non-thermal electrons in a target of a given ambient 
number density. This is a model-independent interpretation (apart 
from the model for the bremsstrahlung cross section). Then it is possible 
to find the ratio of the instantaneous number densities of non-thermal 
electrons ($n_{nt}$) to background density by comparing the non-thermal 
and thermal emission measures. \citet{2013ApJ...771..104F} did this for one flare in 
and found that to explain the HXR observations 
$n_{nt}/n_{background} = 10^{-3}-10^{-4}$ should have $E > 15$ keV. 
But this ratio of course depends on where in the atmosphere the 
HXR emission is produced.

2) One can interpret also the same chromospheric HXR flux 
measurement in a model-dependent way. In the collisional 
thick target model (without re-acceleration) the HXR flux 
measurement can be turned into a flux of electrons arriving 
at the top of the chromosphere of $\approx 10^{19-20}$ electrons cm$^{-2}$ s$^{-1}$ 
e.g. \citep{2011ApJ...739...96K}.
Dividing by the electron speed of $\approx 10^{10}$ cm s$^{-1}$ gives a beam number 
density at the top of the chromosphere of $10^{9-10}$ electrons cm$^{-3}$. 
As the beam slows, $n_{beam}$ (electron beam number density) remains constant so 
$n_{background}$ increases, until the 
beam thermalises. Thus in this model-dependent interpretation the 
fraction number density can vary from $n_{beam}/n_{background} \geq 1$ at the 
top of the chromosphere to $n_{beam}/n_{background} \ll 1$ where the beam stops. 
One should bear in mind also that there is no strong evidence for anisotropic electron 
distributions (beams) in the chromosphere 
(e.g. \citep{2006ApJ...653L.149K,2007A&A...466..705K}).

Generally, the electron energies required to make the non-thermal HXR emission 
in the chromosphere via bremsstrahlung are of the order of tens of keV - speeds of $0.1-0.3 c$. 
This is far less extreme than the above mentioned tens of MeV flare observations.
There are some results that show that HXR 
sources are apparently co-spatial with white-light sources, and produced 
rather low in the atmosphere between 300km and 800km, cf. 
\citep{2012ApJ...753L..26M,2015ApJ...802...19K}.
The photospheric umbral fields can 
go up to $0.2-0.3$ T, and flare sources are also seen in umbrae.
The majority of flare emission, including HXR emission, is 
chromospheric, but there are no observations 
in which we can unambiguously say that the flare does not involve 
the corona in some way. We always see some coronal signature. However, is 
that evidence enough to say that the energy release location is coronal? 
The answer seems uncertain.
Thus there is a good reason  to investigate electron re-acceleration low in the 
chromosphere. For example, the work on the low-altitude HXR and white 
light sources are a motivation, as currently there is no consensus how to get electrons 
accelerated in the corona down to that level, so the implication is 
that they could be accelerated locally.
It should be noted that electron re-acceleration is of relevance not only to
solar flares.
\citet{brunetti07} calculate the acceleration 
of both protons and electrons taking into 
account both transit time damping acceleration and 
non-resonant acceleration by large-scale compressions. 
They find that relativistic electrons can be re-accelerated 
in the  intracluster galactic medium (ICM) up to energies of 
several GeV. 

Section 2 presents the model and results. Section 3 summaries the main findings.

\section{The model and results}

\begin{figure*}
\includegraphics[width=\textwidth]{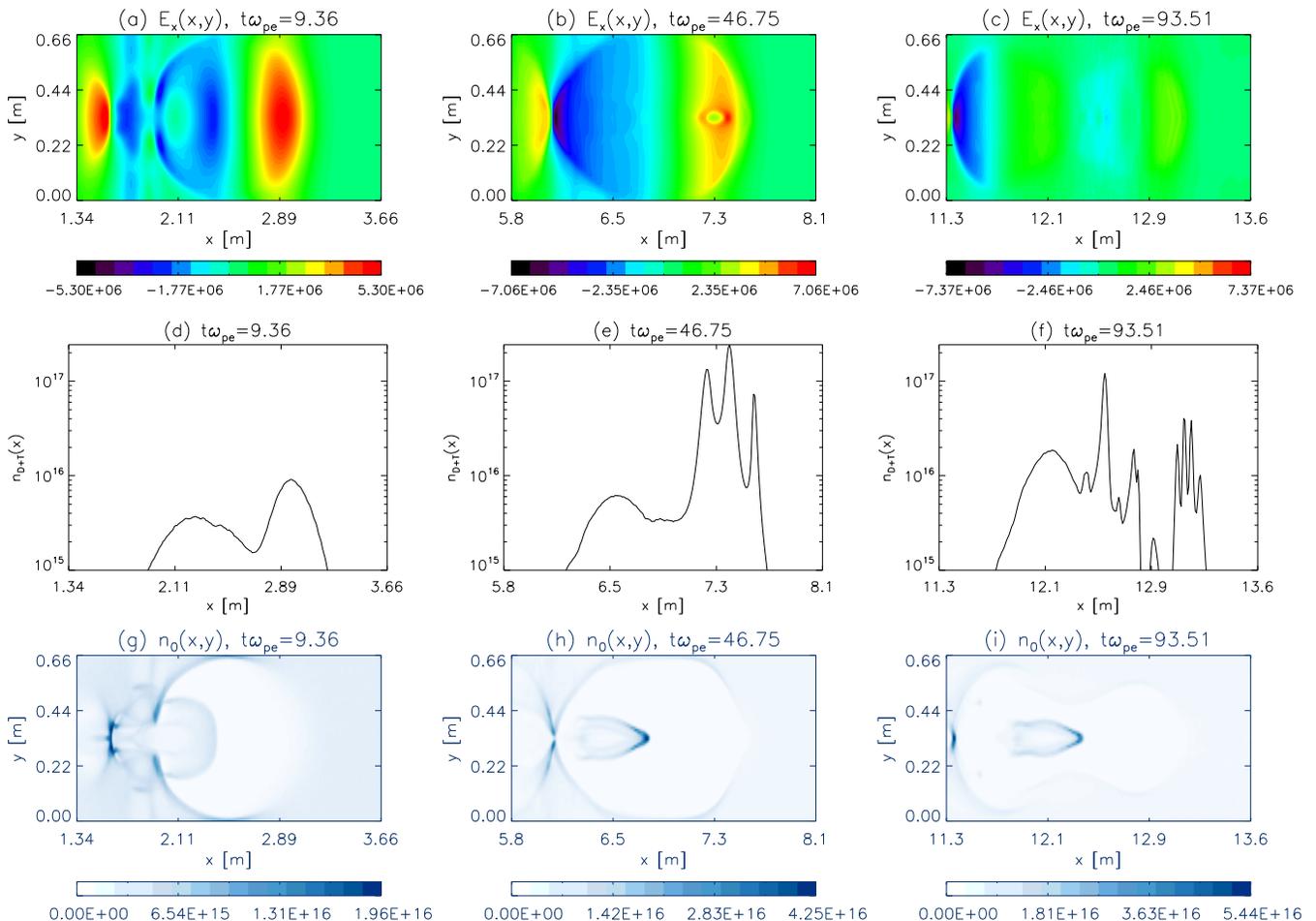}
\caption{(a-c) Contour plots of electric field x-component in (x,y) plane 
(cut through $z=z_{max}/2$) 
at different time instants
corresponding to 1/10th, half and the final simulations times.
(d-f) log normal plot of the sum of driving
and trailing electron bunch number densities at the same times.
(g-i) Contour plots of background electron number density, in units of m$^{-3}$,
in (x,y) plane 
(cut through $z=z_{max}/2$) at the same times.
The fields on color bars are quoted in $V/m$ and time at the top of each panel is
in $ \omega_{pe}$. The data is for solar coronal parameters. See text for details.}
\label{fig1}
\end{figure*}

\begin{figure*}
\includegraphics[width=\textwidth]{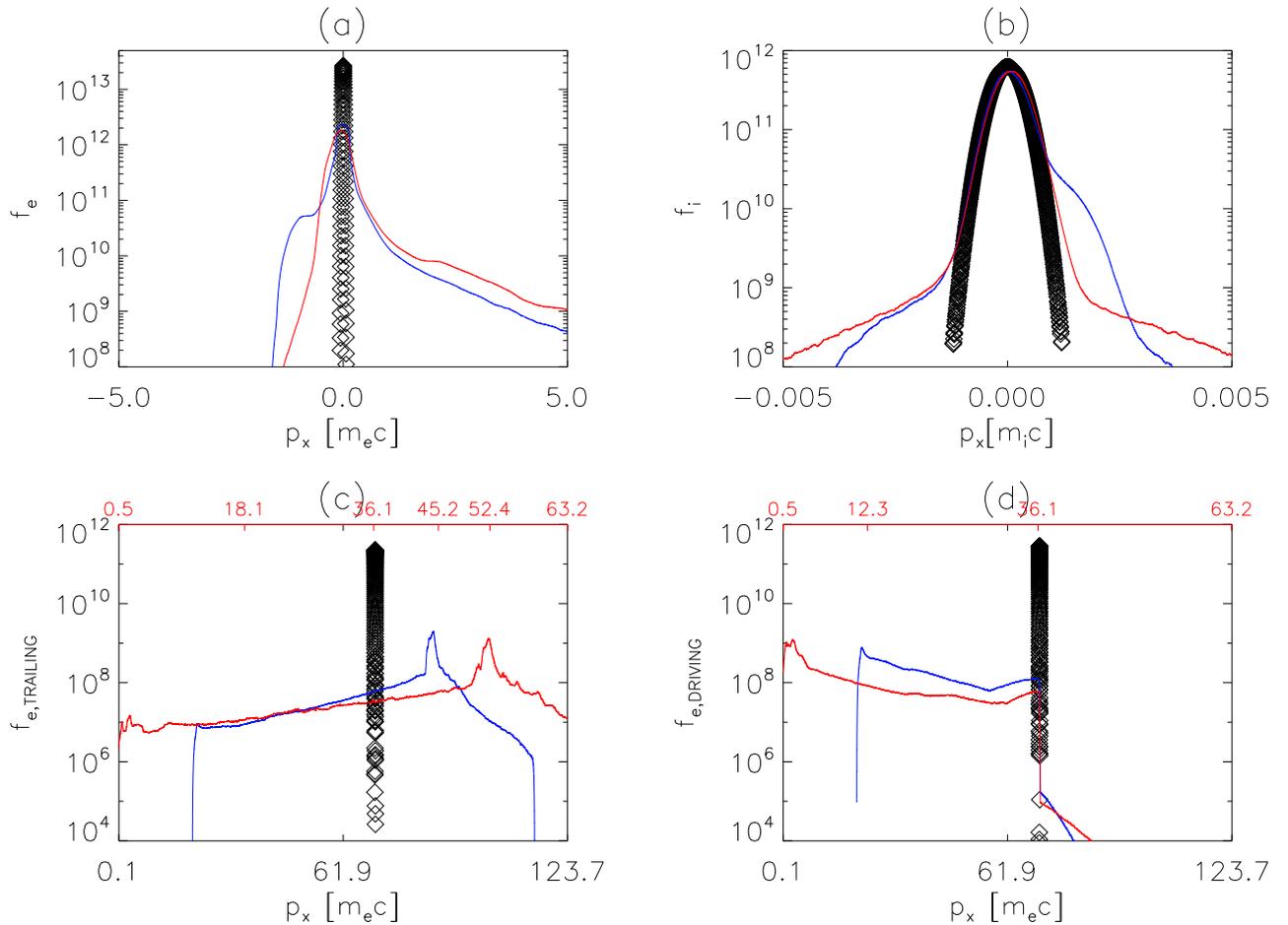}
\caption{Background electron (a), ion (b), trailing (c) and 
driving (d) electron bunch
distribution functions at different times:
open diamonds correspond to $t=0$, while blue and red curves to the 
half and the final simulations times, respectively.
x-axis are momenta quoted in the units of relevant species mass times
speed of light i.e. $[m_e c]$ or $[m_i c]$ as shown on each panel. 
In panels (c) and (d), at the top, the energy is quoted in MeV with red numbers, to aid
eye visualizing of
trailing bunch acceleration and driving bunch deceleration processes.
The data is for solar coronal parameters.}
\label{fig2}
\end{figure*}

\begin{figure} 
\includegraphics[width=\columnwidth]{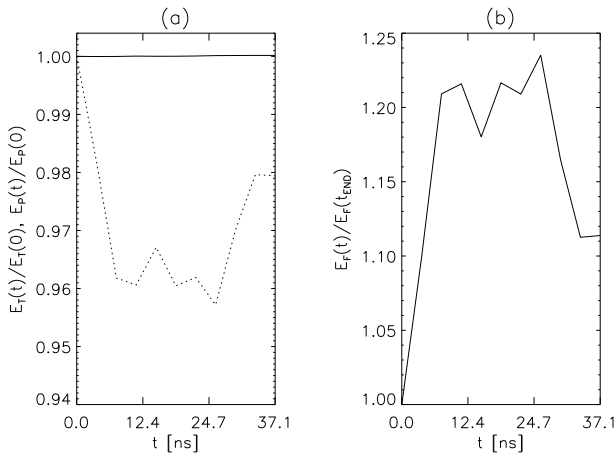}
\caption{In panel (a) solid and dashed curves
are the total (particles plus EM fields) and 
particle energies, normalized on initial values, respectively. 
Panel (b) shows
EM field energy, including background magnetic
field of 0.01 T, normalized on its initial simulation time
value. The data is for solar coronal parameters. The plot is produced
with 10 data points, and time is nano-seconds (ns).}
\label{fig3}
\end{figure}

\begin{figure*}
\includegraphics[width=16.5cm]{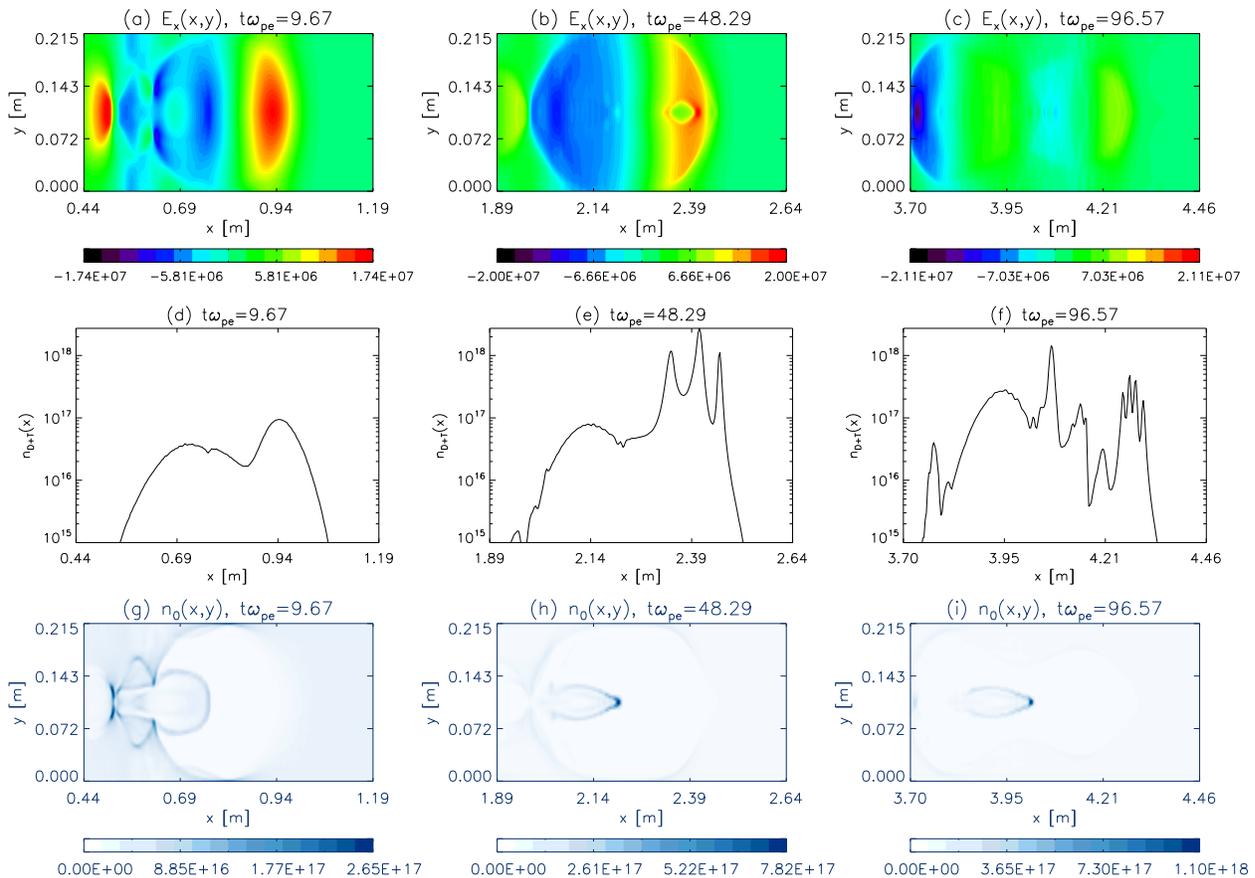}
\caption{As in Fig.\ref{fig1} but for the case of solar chromospheric parameters. See 
text for details.}
\label{fig4}
\end{figure*}

\begin{figure*}
\includegraphics[width=16.5cm]{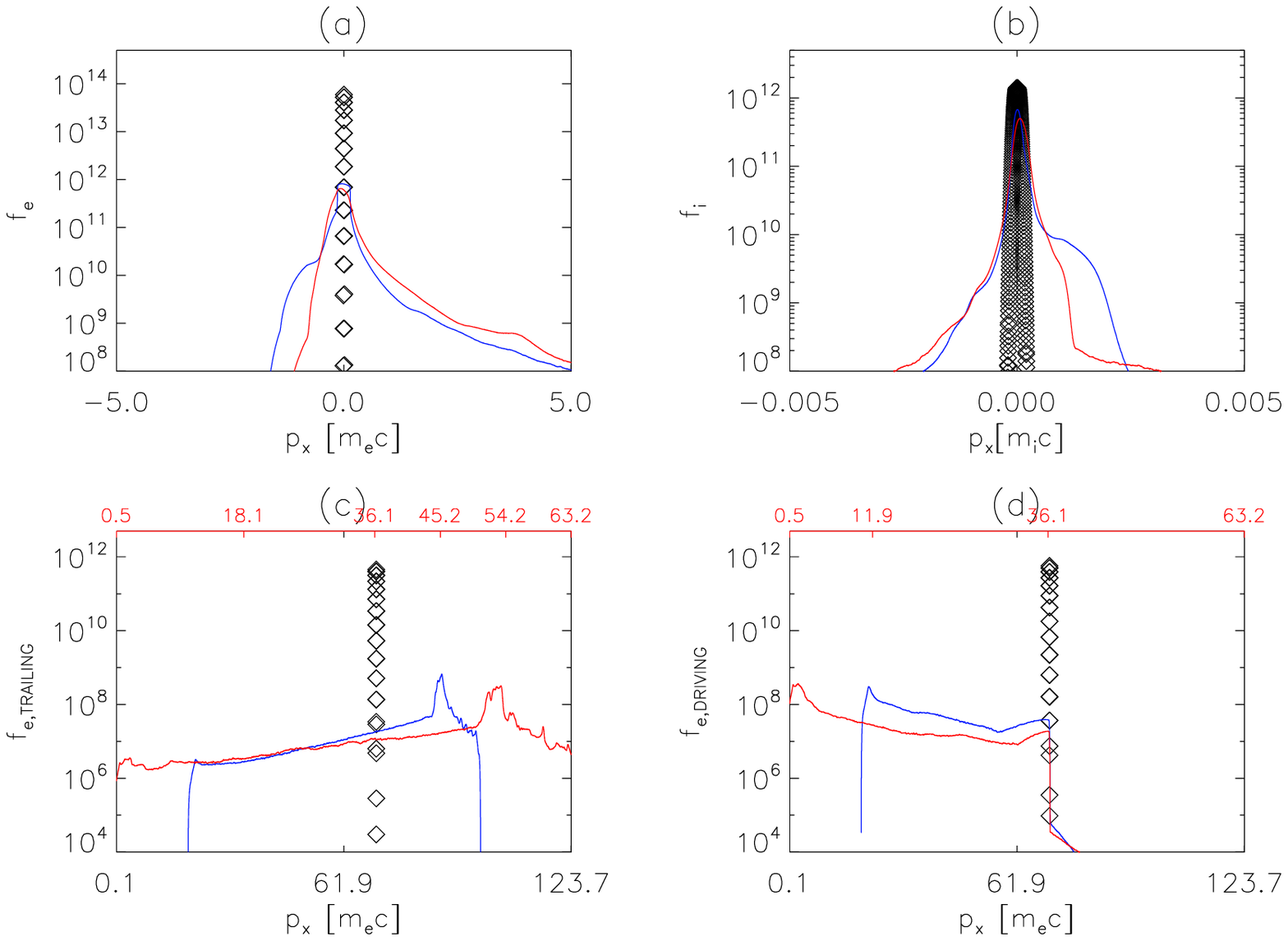}
\caption{As in Fig.\ref{fig2} but for the case of solar chromospheric parameters.}
\label{fig5}
\end{figure*}

\begin{figure} 
\includegraphics[width=\columnwidth]{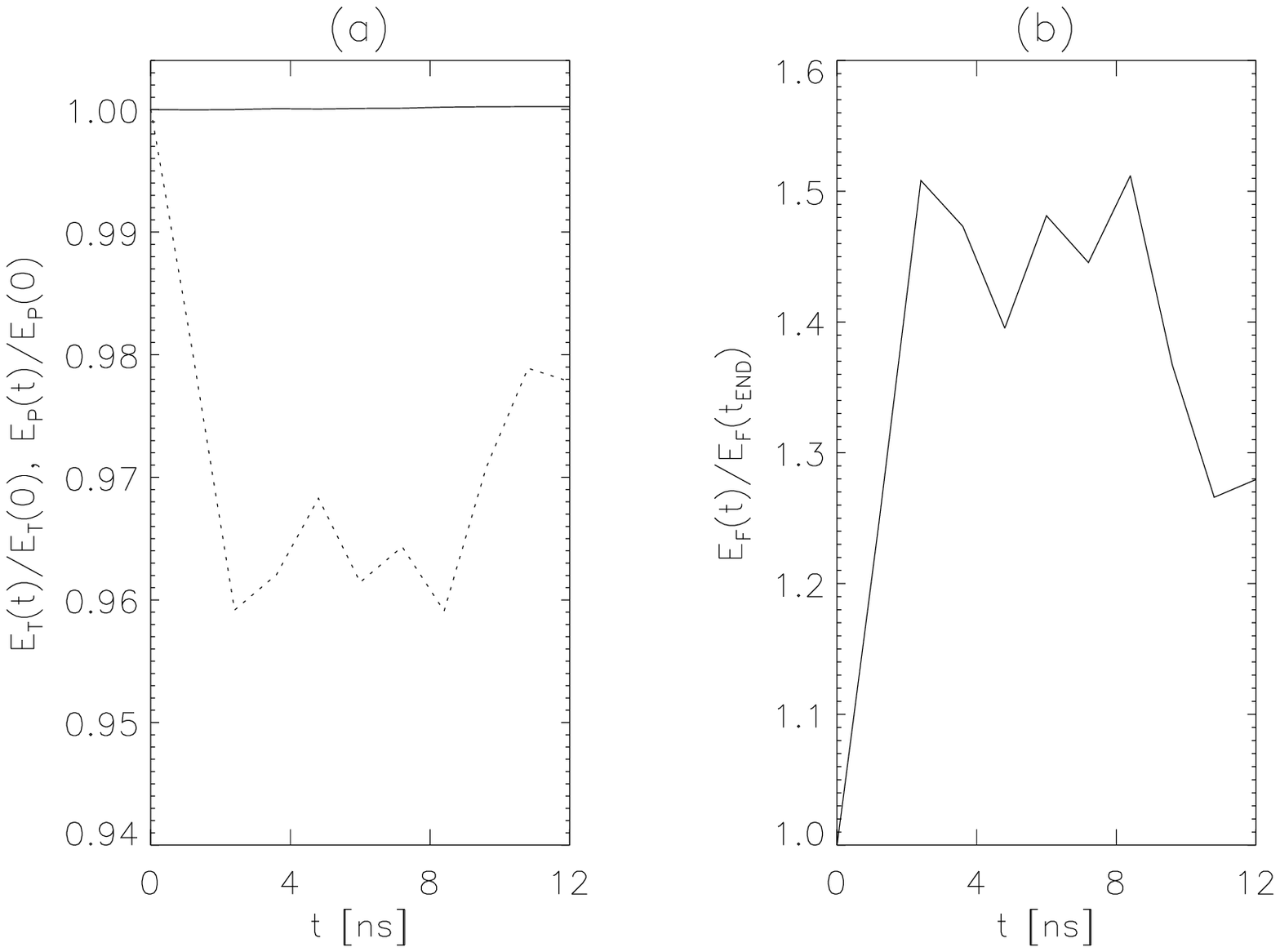}
\caption{As in Fig.\ref{fig3} but for the case of solar chromospheric parameters. Here $B_0=0.02$ T.}
\label{fig6}
\end{figure}

The simulation is carried out using EPOCH, a
fully electromagnetic (EM), relativistic PIC code \cite{a15}.
EPOCH is available for download from
\url{https://cfsa-pmw.warwick.ac.uk}.
The mass ratio in all runs is
$m_i/m_e=1836.153$ and boundary conditions are periodic.

The simulations domain is split into
$n_x \times n_y \times n_z=1500 \times 72 \times 72$
grid cells in x-, y- and z-directions, respectively.
Each grid size is chosen to be Debye length ($\lambda_D$) times appropriate factor ($f$) long.
Here $\lambda_D = v_{th,e}/\omega_{pe}$ denotes the Debye length
with 
$v_{th,e}=\sqrt{k_B T/m_e}$ being electron thermal
speed and $\omega_{pe}$ electron plasma frequency.
This means that as plasma temperature and density is varied
so does the grid size.
In the plasma wake field acceleration
the relevant spatial scale is electron inertial length
$c/\omega_{pe}$. 
We vary factor $f$ such that: 
(i) in the solar coronal run $c/\omega_{pe}$ is
resolved with 12 grid points, i.e.
$(c/\omega_{pe})/\Delta=12.83$, where
$\Delta=f \times \lambda_D$ is the grid size;
(ii) in solar chromosphere run $c/\omega_{pe}$ is
resolved with also 12 grid points, i.e.
$(c/\omega_{pe})/\Delta=12.43$.
This choice seems to provide a reasonable
resolution
since the energy error is small $\approx 0.017-0.025\%$ for the both runs.
Note that $f=6$ in the coronal run and $f=40$ in the chromospheric run.

The trailing and driving electron {\it bunches} have the number
densities as follows:
\begin{eqnarray}
n_{T}(x)= n_0 \times & \exp\left[-\frac{(x-10.0c/\omega_{pe})^2}
{2.0(2.0c/\omega_{pe})^2} \right]  \nonumber \\
& \exp\left[-\frac{(y-y_{max}/2.0)^2}{2.0(c/\omega_{pe})^2}\right]  \nonumber \\
& \exp\left[-\frac{(z-z_{max}/2.0)^2}{2.0(c/\omega_{pe})^2}\right]
\label{e1},
\end{eqnarray}
\begin{eqnarray}
n_{D}(x)=  2.5 n_0 \times & \exp\left[-\frac{(x-16c/\omega_{pe})^2}
{2.0(c/\omega_{pe})^2} \right]  \nonumber \\
& \exp\left[-\frac{(y-y_{max}/2.0)^2}{2.0(c/\omega_{pe})^2}\right]  \nonumber \\
& \exp\left[-\frac{(z-z_{max}/2.0)^2}{2.0(c/\omega_{pe})^2}\right].
\label{e2}
\end{eqnarray}
These expressions imply that trailing bunch is centered
on $10.0c/\omega_{pe}$,
has x-length of $\sigma_x=2.0c/\omega_{pe}$,
while 
driving bunch is 2.5 denser than both the background
and trailing bunch, is centered
on $16.0 c/\omega_{pe}$ and
has x-length of $\sigma_x=c/\omega_{pe}$.
The distance between the trailing and driving
bunches is $6.0c/\omega_{pe}$.
Both electron bunches have
y- and z-lengths of $\sigma_{y,z}=c/\omega_{pe}$
and are centered on $y=y_{max}/2$ and $z=z_{max}/2$.

Both electron bunch initial momenta are set to
$p_x=p_0=\gamma m_e 0.9999c$ kg m s$^{-1}$ 
(note that $p_x/(m_e c)=70.70$, i.e. $\gamma=70.70$),
which corresponds to an initial
energy of $E_0=36.12$ MeV.
There are four plasma species (background electrons and ions, plus
driving and trailing bunches)
present in all numerical simulations.
In the numerical runs there are 
$279,936,000$ particles for each of the four species
i.e. roughly $1.12\times 10^9$ particles
in total.
The three dimensional runs
take about 11 hours on 288 computing cores, using
AMD Interlagos 48-core CPUs with
64 Gb of random access memory and QDR Infiniband.

\subsection{The case of solar coronal plasmas}

In solar coronal parametric run 
uniform density is set to
$n_e=n_i=n_0=2\times10^{15}$ m$^{-3}$    
and temperature to
$T=10^{6}$ K.  Both electron bunch temperatures are also set 
to $T_b=10^{6}$ K. Uniform magnetic field
along x-axis is 0.01 T.

Fig.\ref{fig1} top row shows contour plots of electric field $E_x$ component at three
times. It can be seen that at $t \omega_{pe}=9.36$ the yellow half-ellipse, representing positive 
$E_x \approx 5 \times 10^{6}$ V/m plasma
is followed by blue void of a complex shape with similar amplitude $E_x \approx - {\rm few}
 \times 10^{6}$ V/m.
Panels (a) to (c) show a moving window that has a length of $250 \Delta=250 f \lambda_D$ and width of  
$72 \Delta=72 f \lambda_D$ that
follows the driving and trailing bunches with a speed of $0.9999c$. 
\citet{bulanov16} discuss two effects that impede efficient acceleration:
(i) depletion of either driving laser pulse or electron bunch and
(ii) de-phasing of the trailing electron bunch from the negative electrostatic $E_x$ plasma
wake. Obviously, it is only $E_x <0$ that accelerates the electrons. 
On contrary, $E_x > 0$ results in deceleration of the trailing electron bunch.
Both the electron slippage with respect to the accelerating phase of the wake wave and
the driving bunch / laser pulse energy depletion are both important. 
The de-phasing and depletion lengths are
inversely proportional to the plasma density, and both are of the same order \citep{bulanov16}.
We gather that the both effects are clearly present in panels (a)-(f) of Fig.\ref{fig1}.
In particular we see in panels (a)-(c) that plasma wake strength fades away.
Although we note that on Fig.\ref{fig1}(c) rather compact negative wake near left edge
gains strength to $E_x \approx - 7.4  \times 10^{6}$ V/m.
In panels (d)-(f) we see that initially (panel (d)) driving bunch, the right, taller bump,
is co-spatial with positive (red-yellow) $E_x$ and
trailing bunch, the left, wide-and-short bump, is co-spatial with 
negative (blue) $E_x$. 
This implies that driving bunch will be decelerating,
while trailing bunch accelerating, given the sign of $E_x$.
By the end of simulation time $t \omega_{pe}=93.51$ in panel (f)
the trailing bunch has de-phased from the plasma wake considerably.
From the panels (g)-(i) we gather that initially there was a substantial cavity
created in the background electrons, but it depletes by the end of simulation.

In Fig.\ref{fig2} the details of background electron, ion, trailing and driving electron bunch
distribution functions are quantified  at different times:
open diamonds correspond to $t=0$, while blue and red curves to the 
half and the final simulations times, respectively.
It can be gathered from the plot (panel (a)) that the background electrons
develop non-thermal tails in the direction of
motion of the trailing and driving electron bunches (i.e. positive
x-direction) with values attaining $p_x\approx 5$ $m_e c$.
Ions (panel (b)) initially show beaming in positive $p_x$ direction
(blue curve), but the the end of simulation (red curve)
ions develop non-thermal tails. Note that bulk of the distribution 
does not show significant broadening, only non thermal tails.
Panel (c) demonstrates that by end of simulation
time the trailing bunch gains energy to 52 MeV (red curve), starting from initial 36.1 MeV.
Recall that $\gamma=70.70$ corresponds to the initial
energy of $E_0=36.12$ MeV.
Panel (d) demonstrates that by end of simulation
time the driving bunch loses energy to 0.5 MeV (red curve), starting from initial 36.1 MeV.
This serves as a proof that trailing electron bunch acceleration is
on the expense of driving bunch deceleration. The same conclusion 
can be drawn from the behaviour of different kinds of energies in the next plot.

Panel (a) of Fig.\ref{fig3} shows the behavior of
the total (particles plus EM fields) and 
particle energies, normalized to initial values, respectively. 
The total energy increases due to numerical
heating, but stays within a tolerable value of 0.017 percent, i.e.
$E_T(t)/E_T(0)$ starts from unity and increases to 1.00017.
The particle energy decreases by 4 percent by mid-simulation time and
then bounces back to 0.98 of the initial value.
This points to the fact that the process is intermittent in time.
The reason for such time-transient behaviour is in both 
(i) depletion of electron bunch and
(ii) de-phasing of the trailing electron bunch from the negative electrostatic $E_x$ plasma
wake.
Panel (b) shows
EM field energy normalized to its initial simulation time
value. In can be seen that it experiences time-transient 
increases as the plasma wake is generated and then depleted.

\subsection{The case of solar chromospheric plasmas}

In solar chromosphere parametric run 
uniform density is set to
$n_e=n_i=n_0=2\times10^{16}$ m$^{-3}$    
and temperature to
$T=2.4\times 10^{4}$ K. 
This corresponds to the top of chromosphere
and similar values were used by \citet{tp11}, in a different context.
Both electron bunch temperatures are also set 
to $T_b=2.4\times 10^{4}$ K.
Uniform magnetic field
along x-axis is 0.02 T.

Fig.\ref{fig4} is similar to Fig.\ref{fig1}, but 
for the case of solar chromosphere parameters. 
It can be gathered from panels (a)-(c) of Fig.\ref{fig4} that electrostatic plasma
wake becomes spatially localized (note the different spatial extent on x- and y-axis) 
and, compared to the coronal case, $E_x$ now attains $\sqrt{10}$ larger values of $\pm 1.74 \times 10^{7}$ V/m.
This can be explained by the fact that plasma wake size is prescribed by
the electron inertial length, $c/\omega_{pe}$. Hence, because of the scaling law
$\omega_{pe} \propto \sqrt{n_e}$, larger density, into which
driving bunch plows through, creates more localized and stronger plasma wake.
Panels (d)-(f) show similar bunch de-phasing as in Fig.\ref{fig1},
but now density peak in panel (e) is 10 times higher, as the density in the
chromosphere was chosen to be also 10 times larger.
Panels (g)-(i) also show initial creation and subsequent draining of the
background electron density cavity, except with 20 times larger localised density (note
maximum scale in panel Fig.\ref{fig4}(i)).

Fig.\ref{fig5} is similar to Fig.\ref{fig2} but for the case of solar chromosphere parameters. 
We note in panels (a)-(b) of Fig.\ref{fig5} that 
background electrons and ions have similar response to the injection of
the driving electron bunch, except super-thermal tails are less prominent.
Note also that ion peak at $t=0$, represented by open diamonds in 
panel (b) of  Fig.\ref{fig5}, is narrower than in Fig.\ref{fig2}, as the background
plasma temperature is significantly cooler ($T=2.4\times 10^{4}$ K).
The weaker chromospheric background plasma response can be understood
by its higher density and cooler temperature.
The trailing bunch acceleration and driving bunch
deceleration in the chromospheric case bear the close similarities
to coronal one (as panels (c)-(d) of Fig.\ref{fig5} are similar to Fig.\ref{fig2}).
We should bear in mind that actual domain and thus 
acceleration length are quite different in both cases:
in the solar corona $x_{\rm max} =1500 \Delta= 1500 f \lambda_D =13.888$ m, while
in the chromosphere $x_{\rm max} =1500 \Delta= 1500 f \lambda_D =4.536$ m.
End simulation time in both runs is fixed at $0.8 \times 1500 f \lambda_D /c$, so that the
trailing and driving bunches never reach simulation 
domain boundaries, while traversing its 0.8 length.

Fig.\ref{fig6} is as in Fig.\ref{fig3} but for the case of chromospheric run. 
Here the total energy error is  
0.025 percent (i.e.
$E_T(t)/E_T(0)$ starts from unity and increases to 1.00025). 
This is larger than in coronal run case but still
tolerable. The behaviour of various kinds of energies
is similar to that in Fig.\ref{fig3} in that we still time transient
decrease in the particle and increase in the EM energy.
The notable difference now is that solid curve in the right panel
peaks at 1.5. In the coronal case (Fig.\ref{fig3}) it peaked at 1.2.
Note also that background magnetic fields (used in the normalization) in the both cases are also different.
The stronger peak can be understood by a stronger/denser wake generated
in more dense chromospheric plasma.

\section{conclusions}

This work presents 3D, particle-in-cell, fully electromagnetic simulations of 
electron plasma wake field acceleration applicable to solar atmosphere. 
It was shown that injecting driving and trailing electron bunches
into solar coronal and chromospheric plasmas,
results in electric fields ($-(20-5) \times 10^{6}$ V/m). This leads 
leads to acceleration of the trailing bunch up to 52 MeV, starting from initial 36 MeV.
It is suggested that present results provide one of potentially important
mechanisms for the extreme energetic solar flare 
electron acceleration by means of the
plasma wake field acceleration. It should be noted, however, that
there may exist alternative scenarios. For example, it is well known that the
magnetic reconnection of solar coronal fields (see e.g., \citet{2002tt}) 
can give rise to strong field-aligned electric fields. Such fields could also
give rise to a mechanism of accelerating electrons to high energies. Similar 
examples may
be found in the geomagnetic field implication of reconnection (e.g. \citet{wag80}). 
Another examples may be found in the direct current (DC) field 
acceleration due to reconnection
fields including laboratory plasmas (e.g. \citet{leb79,leb82}).

%
%
%

\begin{acknowledgments}
This research utilized Queen Mary University of London's (QMUL) 
MidPlus computational facilities,       
supported by QMUL Research-IT and funded by UK EPSRC grant EP/K000128/1.
\end{acknowledgments}


\end{document}